# Solitary Wave Solutions of KP equation, Cylindrical KP Equation and Spherical KP Equation[∗]

LI Xiang-Zheng[1], ZHANG Jin-Liang[1] & WANG Ming-Liang[1,2]

1. School of Mathematics & Statistcs, Henan University of Science & Technology, Luoyang, 471023, PR China

2. School of Mathematics & Statistcs, Lanzhou University, Lanzhou, 730000, PR China

***Abstract*** *Three (2+1)-dimensional equations, they are KP equation, cylindrical KP equation and spherical KP equation, have been reduced to the same KdV equation by different transformation of variables respectively. Since the single solitary wave solution and 2-solitary wave solution of the KdV equation have been known already, substituting the solutions of the KdV equation into the corresponding transformation of variables respectively, the single and 2-solitary wave solutions of the three (2+1)-dimensional equations can be obtained successfully.*

## 1. Introduction

Many famous nonlinear evolution equations such as Korteweg-de Vries (KdV), Modified KdV (mKdV), Kadomstev-Perviashvili (KP), Coupled KP and Zakharov-Kuznetsov (ZK) have been obtained in nonlinear propagation of dust-acoustic wave, especially, the dust-acoustic solitary wave (DASW) in space and laboratory plasma[1]. It is known that the transverse perturbations always exist in the higher-dimensional system. Anisotropy is introduced into the system and the wave structure and stability are modified by the transverse perturbation. Recent theoretical studies for ion-acoustic/dust-acoustic waves show that the properties of solitary waves in bounded nonplanar cylindrical/spherical geometry differ from that in unbounded planar geometry. A dissipative cylindrical/spherical KdV is obtained by using the standard reductive perturbation method in Ref. [2]. The cylindrical KP equation (CKP) has been introduced by Johnson[3,4] to describe surface wave in a shallow incompressible fluid. A spherical KP (SKP) equation is obtained by using the standard reductive perturbation method[5].

We consider the classical KP equation in the form

$$(u_t + 6uu_x + u_{xxx})_x + su_{yy} = 0, \qquad (1)$$

where $s$ is a constant. The KP equation is also derived using reductive perturbation method in superthermal dusty plasma and the steady state solution has been given[6].

∗E-mail: 57546388@qq.com

∗ The project supported in part by the National Natural Science Foundation of China(Grant No. 11301153) and The Doctoral Foundation of Henan University of Science and Technology(Grant No. 09001562) and The Science and Technology Innovation Platform of Henan University of Science and Technology(Grant No.2015XPT001) .

The cylindrical KP equation (CKP)

$$\left(u_t + 6uu_x + u_{xxx} + \tfrac{1}{2t}u\right)_x + \tfrac{3\alpha^2}{t^2}u_{yy} = 0, \tag{2}$$

where $\alpha$ is a constant, has also been investigated to obtain decay mode solutions by means of Hirota method and simplified homogeneous balance method respectively[7-8].

We also consider the classical spherical KP equation (SKP) in the form

$$\left(u_t + 6uu_x + u_{xxx} + \tfrac{1}{t}u\right)_x + \tfrac{1}{2v_0 t^2}\left(u_{yy} + \tfrac{1}{y}u_y\right) = 0, \tag{3}$$

where $v_0$ is a constant. An exact solitary wave solution of SKP, that demonstrates the amplitude and wave velocity of solitary wave are uniquely determined by the parameters of the system and only depending on the initial conditions, has been obtained in Ref. [5].

The KdV equation has been researched by many authors. The multi-soliton solutions of the KdV equation with general variable coefficients have been completely obtained by homogeneous balance principle[9,10]. Some solutions, which possess movable singular points while their energies are only redistributed without dissipation, of KdV equation have been obtained through the modified bilinear Bäcklund transformation[11]. A n-soliton solution with the bell shape has been obtained in Ref. [12], whose stationary height is an arbitrary constant $c$. The rational solutions, solitary wave solutions, triangular periodic solutions, Jacobi periodic wave solutions and implicit function solutions of KdV equation have been constructed by means of an extended algebraic method[13]. If a transformation of variables between the KdV equation and other equation can be constructed, the results of KdV equation above can be used directly.

In the present paper we aim to find solitary wave solutions of KP Eq.(1), CKP Eq.(2) and SKP Eq.(3). The paper is organized as follows: In section 2, the KP Eq.(1), CKP Eq.(2) and SKP Eq.(3) are reduced to the same KdV equation by different transformation of variables respectively. In section 3, the single solitary wave solutions and 2-solitary wave solutions of KP Eq.(1), CKP Eq.(2) and SKP Eq. (3) can be obtained in terms of the corresponding transformation of variables respectively, since the solutions of the KdV equation have been known already. In section 4, some conclusions are made.

## 2. Reduction of KP, CKP and SKP
### 2.1 Reduction of KP

In Eq. (1) we assume that

$$u(x,y,t) = w(\xi,t),\ \xi = x + q_1(y,t), \tag{4}$$

where $q_1 = q_1(y,t)$ is a function to be determined later. Substituting (4) into (1), yields a equation as follows

$$\tfrac{\partial}{\partial \xi}\left(w_t + 6ww_\xi + w_{\xi\xi\xi}\right) + sq_{1yy}w_\xi + (sq_{1y}^2 + q_{1t})w_{\xi\xi} = 0. \tag{5}$$

Setting the coefficients of $w_\xi$ and $w_{\xi\xi}$ to zero, yields

$$sq_{1yy} = 0,\ \ sq_{1y}^2 + q_{1t} = 0. \tag{6}$$

The system (6) admits the following solution:

$$q_1(y,t) = \lambda y - s\lambda^2 t, \tag{7}$$

where $\lambda$ is a nonzero arbitrary constant. Using (7) the expression (4) becomes

$$u = w(\xi,t), \xi = x + \lambda y - s\lambda^2 t. \tag{8}$$

And after integrating (5) with respect to $\xi$ once and taking the constant of integration to zero, equation (5) becomes the classical KdV equation for $w = w(\xi,t)$

$$w_t + 6ww_\xi + w_{\xi\xi\xi} = 0. \tag{9}$$

From the discussion above, we come to the conclusion that the KP Eq.(1) for $u = u(x, y, t)$ is reduced to the KdV Eq.(9) for $w = w(\xi, t)$ by using the transformation of variables (8), if $w(\xi, t)$ is a solution of KdV Eq.(9), substituting it into (8), then we have the exact solution of the KP Eq.(1).

### 2.2 Reduction of CKP

In Eq.(2) we assume that

$$u(x, y, t) = w(\xi, t), \xi = x + q_2(y, t), \tag{10}$$

where $q_2 = q_2(y,t)$ is a function to be determined later. Substituting (10) into (2), yields a equation as follows

$$\frac{\partial}{\partial \xi}\left(w_t + 6ww_\xi + w_{\xi\xi\xi}\right) + \frac{t+6\alpha^2 q_{2yy}}{2t^2} w_\xi + (\frac{3\alpha^2}{t^2} q_{2y}^2 + q_{2t})w_{\xi\xi} = 0. \tag{11}$$

Setting the coefficients of $w_\xi$ and $w_{\xi\xi}$ to zero, yields

$$t + 6\alpha^2 q_{2yy} = 0, \frac{3\alpha^2}{t^2} q_{2y}^2 + q_{2t} = 0. \tag{12}$$

The system (12) admits a solution:

$$q_2(y,t) = -\frac{1}{12\alpha^2} y^2 t + yt - 3\alpha^2 t. \tag{13}$$

Using (13) the expression (10) becomes

$$u = w(\xi,t), \xi = x - \frac{1}{12\alpha^2} y^2 t + yt - 3\alpha^2 t. \tag{14}$$

And after integrating (11) with respect to $\xi$ once and taking the constant of integration to zero, the equation (11) becomes the classical KdV Eq.(9) for $w(\xi,t)$.

From the discussion above, we come to the conclusion that the CKP Eq.(2) for $u = u(x, y, t)$

is reduced to the KdV Eq.(9) for $w = w(\xi,t)$ by using the transformation of variables (14), if $w(\xi,t)$ is a solution of KdV Eq.(9), substituting it into (14), then we have the exact solution of CKP Eq.(2)

**2.3 Reduction of SKP**

In Eq.(3), we assume that

$$u(x,y,t) = w(\xi,t), \xi = x + q_3(y,t), \tag{15}$$

where $q_3 = q_3(y,t)$ is a function to be determined later. Substituting (15) into (3), yields a equation as follows

$$\frac{\partial}{\partial \xi}\left(w_t + 6ww_\xi + w_{\xi\xi\xi}\right) + \frac{q_{3y} + y(2v_0 t + q_{3yy})}{2v_0 t^2 y} w_\xi + \left(\frac{1}{2v_0 t^2} q_{3y}^2 + q_{3t}\right) w_{\xi\xi} = 0. \tag{16}$$

Setting the coefficients of $w_\xi$ and $w_{\xi\xi}$ to zero, yields

$$q_{3y} + y(2v_0 t + q_{3yy}) = 0, \frac{1}{2v_0 t^2} q_{3y}^2 + q_{3t} = 0. \tag{17}$$

The system (17) admits a solution:

$$q_3(y,t) = -\frac{v_0}{2} y^2 t + \mu, \tag{18}$$

where $\mu$ is a nonzero arbitrary constant. Using (18) the expression (15) becomes

$$u = w(\xi,t), \xi = x - \frac{v_0}{2} y^2 t + \mu. \tag{19}$$

and after integrating the equation (16) once, taking the constant of integration to zero, the equation (16) becomes the classical KdV Eq.(9).

From the discussion above, we come to the conclusion that the SKP Eq.(3) for $u = u(x,y,t)$ is reduced to the KdV Eq.(9) for $w = w(\xi,t)$ by using the transformation of variables (19), if $w(\xi,t)$ is a solution of KdV Eq.(9), and substituting it into (19), we have the exact solution of the SKP Eq.(3).

**3. Solitary wave solutions of KP,CKP and SKP**

In previous section, the KP Eq.(1),CKP Eq.(2) and SKP Eq.(3) have been reduced into the same classical KdV Eq.(9) by the transformation (8),(14) and (19), respectively. The KdV Eq.(9) is of physically importance and its solutions have been known for many researchers, for instance, according to Ref. [9], the KdV Eq.(9) has single solitary wave

$$w(\xi,t) = 2k^2 \frac{e^\eta}{(1+e^\eta)^2}, \quad \eta = k\xi - k^3 t + x_0, \tag{20}$$

where $k$ and $x_0$ are arbitrary parameters. And KdV Eq.(9) also has 2-soliton solution

$$w(\xi,t) = 2\frac{k_1^2 e^{\eta_1} + k_2^2 e^{\eta_2} + 2(k_1-k_2)^2 e^{\eta_1+\eta_2} + a_{12}\left(k_2^2 e^{2\eta_1+\eta_2} + k_1^2 e^{\eta_1+2\eta_2}\right)}{\left(1+e^{\eta_1}+e^{\eta_2}+a_{12}e^{\eta_1+\eta_2}\right)^2}, \tag{21}$$

where $k_i$ and $x_i$ are arbitrary parameters, $\eta_i = k_i\xi - k_i^3 t + x_i$, $(i=1,2)$, $a_{12} = \frac{(k_1-k_2)^2}{(k_1+k_2)^2}$.

**3.1 solitary wave solutions of KP**

Substituting (20) into (8) we have the single solitary wave solution for KP Eq. (1), which is expressed by

$$u(x,y,t) = 2k^2 \frac{e^\eta}{(1+e^\eta)^2}, \quad \eta = k(x+\lambda y) - (ks\lambda^2 + k^3)t + x_0, \tag{22}$$

where $k$ and $x_0$ are arbitrary parameters.

Substituting (21) into (8) we have the 2-solitary solution for KP Eq.(1), which is expressed by

$$u(x,y,t) = 2\frac{k_1^2 e^{\eta_1} + k_2^2 e^{\eta_2} + 2(k_1-k_2)^2 e^{\eta_1+\eta_2} + a_{12}\left(k_2^2 e^{2\eta_1+\eta_2} + k_1^2 e^{\eta_1+2\eta_2}\right)}{\left(1+e^{\eta_1}+e^{\eta_2}+a_{12}e^{\eta_1+\eta_2}\right)^2}, \tag{23}$$

where $k_i$ and $x_i$ are arbitrary parameters, $\eta_i = k_i(x+\lambda y) - (k_i s\lambda^2 + k_i^3)t + x_i$, $(i=1,2)$, $a_{12} = \frac{(k_1-k_2)^2}{(k_1+k_2)^2}$.

**3.2 solitary wave solutions of CKP**

Substituting (20) into (14) we have the single solitary wave solution for CKP Eq.(2), which is expressed by

$$u(x,y,t) = 2k^2 \frac{e^\eta}{(1+e^\eta)^2}, \quad \eta = k[x - (\frac{1}{12\alpha^2} y^2 - y + 3\alpha^2 + k^2)t] + x_0, \tag{24}$$

where $k$ and $x_0$ are arbitrary parameters. The solution (24) is shown in Figs.1-4 with $k = 0.2$, $x_0 = 3$, $\alpha = 1$.

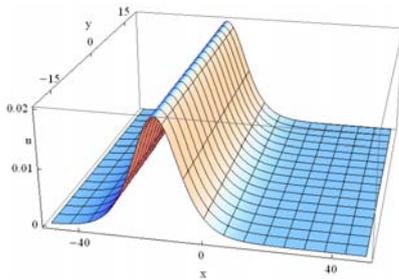 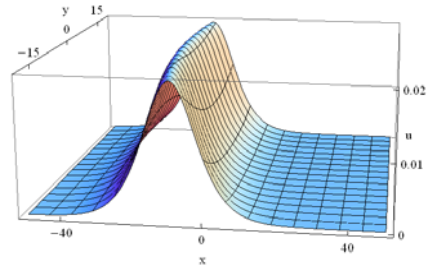

Fig.1. Plots of solution (24) with $t=0$           Fig.2. Plots of solution (24) with $t=0.1$

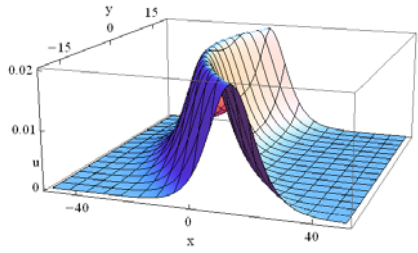
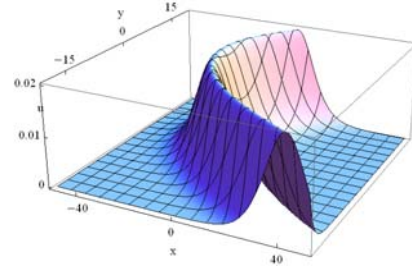

Fig.3. Plots of solution (24) with $t = 0.5$          Fig.4. Plots of solution (24) with $t = 1$

Substituting (21) into (14) we have the 2-solitary solution for CKP Eq.(2), which is expressed by

$$u(x,y,t) = 2\frac{k_1^2 e^{\eta_1} + k_2^2 e^{\eta_2} + 2(k_1-k_2)^2 e^{\eta_1+\eta_2} + a_{12}\left(k_2^2 e^{2\eta_1+\eta_2} + k_1^2 e^{\eta_1+2\eta_2}\right)}{\left(1+e^{\eta_1}+e^{\eta_2}+a_{12}e^{\eta_1+\eta_2}\right)^2}, \quad (25)$$

where $k_i$ and $x_i$ are arbitrary parameters, $a_{12} = \frac{(k_1-k_2)^2}{(k_1+k_2)^2}$, $\eta_i = k_i[x - (\frac{1}{12\alpha^2}y^2 - y + 3\alpha^2 + k_i^2)t] + x_i$, $(i=1,2)$. The solution (25) is shown in Figs.5-8 with $\alpha = 1$, $x_1 = -2$, $x_2 = 2$, $k_1 = 0.2$, $k_2 = 0.3$.

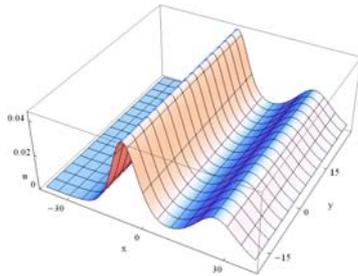
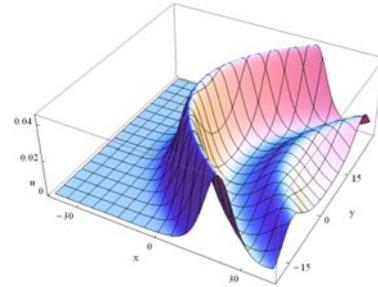

Fig.5. Plots of solution (25) with $t = 0$          Fig.6. Plots of solution (25) with $t = 0.5$

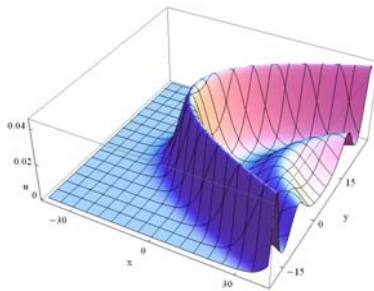
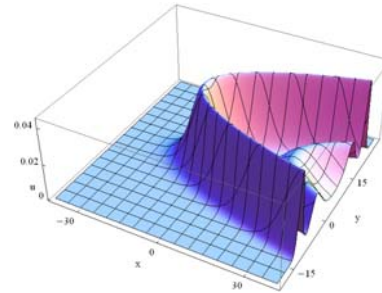

Fig.7. Plots of solution (25) with $t = 1$          Fig.8. Plots of solution (25) with $t = 2$

### 3.3 solitary wave solutions of SKP

Substituting (20) into (19) we have the single solitary wave solution for SKP Eq. (3), which is expressed by

$$u(x,y,t) = 2k^2 \frac{e^\eta}{(1+e^\eta)^2}, \quad \eta = k[x - (\frac{v_0}{2}y^2 + k^2)t + \mu] + x_0, \quad (26)$$

where $k$ and $x_0$ are arbitrary parameters.

Substituting (21) into (19) we have the 2-solitary solution for SKP Eq. (3), which is expressed by

$$u(x,y,t) = 2\frac{k_1^2 e^{\eta_1} + k_2^2 e^{\eta_2} + 2(k_1-k_2)^2 e^{\eta_1+\eta_2} + a_{12}\left(k_2^2 e^{2\eta_1+\eta_2} + k_1^2 e^{\eta_1+2\eta_2}\right)}{\left(1+e^{\eta_1}+e^{\eta_2}+a_{12}e^{\eta_1+\eta_2}\right)^2}, \qquad (27)$$

$$\eta_i = k_i[x - (\tfrac{v_0}{2}y^2 + k_i^2)t] + x_i,$$

where $k_i$ and $x_i$ are arbitrary parameters, $a_{12} = \frac{(k_1-k_2)^2}{(k_1+k_2)^2}$, $(i=1,2)$.

## 4. Conclusion

In this paper, by making corresponding transformation of variables, the KP equation, cylindrical KP equation and spherical KP equation are all reduced to the same classical KdV equation, which can be solved by using homogeneous balance method[9,10] to obtain single solitary wave solution and 2-soliton solution. Substitutiong the solitary solutions of the KdV equation into the corresponding transformation of variables, we have the solitary wave solutions of the KP equation, cylindrical KP equation and spherical KP equation, respectively. It is interesting to research CKP equation and SKP equation but avoid the singularity point analysis when $t=0$. The analysis in the present paper may be extended to other works to make further progress.

**Acknowledgments**

The authors are very grateful to the referees for their invaluable comments.

**References**
[1] H. Y. Wang and K.B. Zhang, *J. Sichuan Normal Univ.* **36** (2013) 911. (in Chinese)
[2] J. K. Xue, *Phys. Lett. A* **322** (2004) 225.
[3] R. S. Johnson, *J. Fluid Mech.* **97** (1980) 701.
[4] R. S. Johnson, *A modern introduction to the mathematical theory of water waves,* Cambridge University Press, Cambridge (1997).
[5] J. K. Xue, *Phys. Lett. A* **314** (2003) 479.
[6] N. S. Saini, Nimardeep Kaur and T. S. Gill, *Advan. Space Res.* **55** (2015) 2873.
[7] S.F. Deng, *Appl. Math. Comput.* **218** (2012) 5974.
[8] M. L. Wang, J. L. Zhang and X. Z. Li, *Appl. Math. Lett.* **62** (2016) 29.
[9] M. L. Wang and Y. M. Wang, *Phys. Lett. A* **287** (2001) 211.
[10] W. P. Zhong, R. H. Xie, M. Belić, N. Petrović, G. Chen, and L. Yi, *Phys. Rev. A* **78**(2008)023821.
[11] Z. Y. Chen, J. B. Bi, D. Y. Chen, *Commun. Theor. Phys.* **41**(2004)397.
[12] F. K. Guo, Y. F. Zhang, *Commun. Theor. Phys.* **46**(2006)577.
[13] X. L. Yang, J. S. Tang, *Commun. Theor. Phys.* **48**(2007)1.

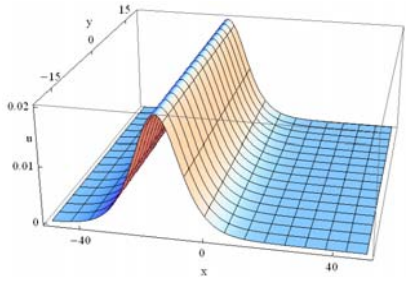
Fig.1. Plots of solution (24) with $t = 0$

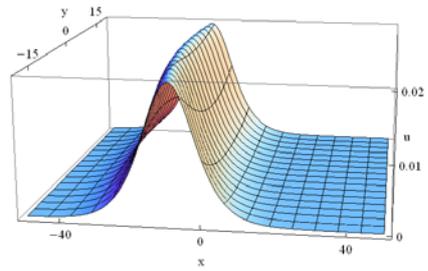
Fig.2. Plots of solution (24) with $t = 0.1$

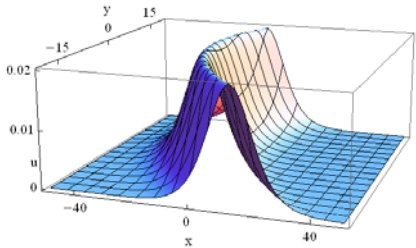
Fig.3. Plots of solution (24) with $t = 0.5$

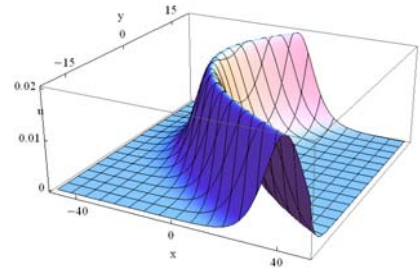
Fig.4. Plots of solution (24) with $t = 1$

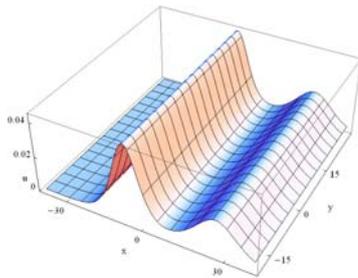
Fig.5. Plots of solution (25) with $t = 0$

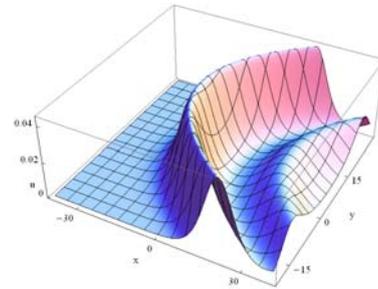
Fig.6. Plots of solution (25) with $t = 0.5$

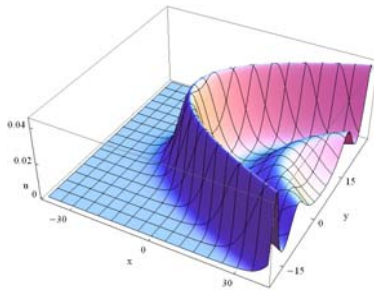
Fig.7. Plots of solution (25) with $t = 1$

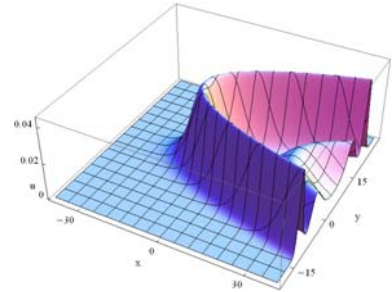
Fig.8. Plots of solution (25) with $t = 2$